# Real-Space Investigation of the Charge Density Wave in VTe$_2$ Monolayer with Rotational and Mirror Symmetries Broken

Guangyao Miao[1,2], Siwei Xue[1,2], Bo Li[1,2], Zijian Lin[1,2], Bing Liu[1,2], Xuetao Zhu[1,2,3]*, Weihua Wang[1]*, and Jiandong Guo[1,2,3,4]

[1] *Beijing National Laboratory for Condensed Matter Physics and Institute of Physics, Chinese Academy of Sciences, Beijing 100190, China*

[2] *School of Physical Sciences, University of Chinese Academy of Sciences, Beijing 100049, China*

[3] *Songshan Lake Materials Laboratory, Dongguan, Guangdong 523808, China*

[4] *Beijing Acadamy of Quantum Information Sciences, Beijing 100193, China*

*Emails: xtzhu@iphy.ac.cn; weihuawang@iphy.ac.cn.

**Abstract**

Recently the charge density wave (CDW) in vanadium dichalcogenides have attracted increasing research interests, but a real-space investigation on the symmetry breaking of the CDW state in VTe$_2$ monolayer is still lacking. We have investigated the CDW of VTe$_2$ monolayer by low energy electron diffraction (LEED) and scanning tunneling microscope (STM). While the LEED experiments revealed a (4×4) CDW transition at 192±2 K, our low-temperature STM experiments resolved the (4×4) lattice distortions and charge-density modulation in real space, and further unveiled a 1D modulation that breaks the three-fold rotational and mirror symmetries in the CDW state. In accordance with the CDW state at low temperature, a CDW gap of 12 meV was detected by scanning tunneling spectroscopy (STS) at 4.9 K. Our work provides real-space evidence on the symmetry breaking of the (4×4) CDW state in VTe$_2$ monolayer, and implies there is a certain mechanism, beyond the conventional Fermi surface nesting or the q-dependent electron-phonon coupling, is responsible for the formation of CDW state in VTe$_2$ monolayer.

Charge density wave (CDW) is the ground state of many low-dimensional materials, germane to various physical phenomena from peculiar transport properties to the mechanism of superconductivity [1-5]. The layered transition metal dichalcogenides (TMDs) are the prototypical CDW systems that have been mostly studied since 1970s [6]. Recent breakthroughs in the synthesis techniques have inspired the studies of CDW in TMD monolayers, which shows distinctive properties from the layered TMD bulk materials due to quantum confinement and reduced screening [7]. For example, monolayers of $1T$-TiSe$_2$ and $1T$-VSe$_2$ have higher CDW transition temperatures than their bulk counterparts [8-11], and $1T$-VSe$_2$ monolayer exhibits varied superstructures different from the layers of the bulk [10-12]. These studies not only deepen our understanding of two-dimensional (2D) materials, but also enhance our knowledge of CDW from the usual quasi-2D layered materials to the real 2D limit.

The formation of CDW is a spontaneous symmetry-breaking process involving charge density modulation accompanied by periodic lattice distortions [5, 6, 13]. Experimental determination of the lattice periodicity and symmetry differences between the normal state and the CDW state is always the first priority for any CDW study. Different from the diffraction methods that reveal the averaged superstructures of the CDW state at macroscale in the reciprocal space, scanning tunneling microscope (STM) provides valuable information about both the distortions of atomic structures and distribution of electronic states in real space, proven to be a powerful tool to study CDWs in low-dimensional systems [14-19]. In most cases, the formation of superstructures in a CDW state only breaks the translational symmetry and sometimes the mirror symmetry of the normal state, with the rotational symmetry reserved [8, 12, 15, 19-22]. Recently, a CDW state with ($\sqrt{3} \times \sqrt{7}$) superstructures has been reported in monolayer VSe$_2$, which breaks both the translational and rotational symmetries of the normal state [10, 11]. However, a similar system of vanadium dichalcogenides, monolayer VTe$_2$, was reported to have a (4×4) CDW state with the three-fold rotational symmetry reserved as shown by spectroscopic methods in the reciprocal space [23]. This seems to be very different from the case of VSe$_2$ monolayer with the breaking of rotational symmetry in the CDW state, and thus calls for a detailed investigation in real-space.

Here we investigated the (4×4) CDW state in VTe$_2$ monolayers by low-temperature STM and low energy electron diffraction (LEED) in detail, and report that in addition to the translational symmetry, both the three-fold rotational and the mirror symmetries are actually broken in the CDW state. Our work implies the CDW in VTe$_2$ monolayer is possibly formed due to a mechanism beyond the conventional Fermi surface nesting (FSN) or the q-dependent electron-phonon coupling (EPC).

The bulk VTe$_2$ has a layered structure of CdI$_2$-type with space group P$\bar{3}m1$(164) (1$T$-phase), and changes to a monoclinic phase below 482 K [24]. Recently VTe$_2$ thin films with the thickness from monolayer to 60 nm have been synthesized by several groups, and found to take the 1$T$ structure [25-27]. Each 1$T$-VTe$_2$ layer is composed of VTe$_6$ octahedra by sharing their edges, and in this way, a layer of V atoms is sandwiched between two layers of Te atoms in triangular lattices [Fig. 1(a)]. In our experiments, the VTe$_2$ monolayers are grown on graphene/SiC substrates by molecular-beam epitaxy, and investigated by *in-situ* STM and *ex-situ* LEED [28]. Figure 1(b) shows the large-scale STM image of a VTe$_2$ monolayer on graphene/SiC substrate. The VTe$_2$ monolayer spreads over two adjacent graphene terraces, manifesting its 2D characteristic.

Figure 1(c) shows the high-resolution STM image of a VTe$_2$ monolayer scanned at 290 K. The protrusions in the image are attributed to Te atoms of the top sublayer. One can see that the VTe$_2$ monolayer has three-fold rotational and mirror symmetries and a (1×1) structure with lattice constants *a*=*b*=0.37 nm [Fig. 1(c)], which agrees well with the previous reports [23, 25, 26]. The high-resolution STM image scanned at 78 K reveals that the VTe$_2$ monolayer has a (4×4) superstructure, as indicated by the unit cell defined by *a'* and *b'* in the image. The measured period of the unit cell is *a'*=*b'*=1.48 ± 0.09 nm, which exactly gives *a'*=*b'*=4*a*=4*b*. The structure change between Fig. 1(c) and Fig. 1(d) implies the existence of a (4×4) CDW transition at some temperature in the range from 290 K to 78 K [23]. This is corroborated by the contrast inversion between the occupied- and empty-state STM images at 78 K [28], which is a characteristic feature of the CDW state in STM measurements [14, 17, 18].

To reveal the CDW transition temperature, we performed LEED measurements on VTe$_2$ monolayers grown on graphene/SiC substrate [28]. Figures 2(a)~2(e) show a series of LEED patterns obtained at various temperatures elevated from 35 K to 191 K. Two sets of diffraction spots, the outmost six sharp spots and inner six blurred ones, are observed all through the temperature range. The outmost spots are given by the graphene substrate, as confirmed by the LEED pattern measured on bare graphene/SiC substrate [Fig. 2(f)]. The inner spots are attributed to the 1$T$-VTe$_2$ lattice with an in-plane lattice constant of 0.35 ± 0.03 nm, in agreement with our STM results. The broadening of the 1$T$-VTe$_2$ spots is probably given by different VTe$_2$ monolayer domains with slightly varied orientations.

In addition to the diffraction spots given by graphene and 1$T$-VTe$_2$, fractional spots are observed in LEED patterns below 190 K, as marked by the arrows in Fig. 2(a). The fractional spots indicate that a (4×4) superstructure is formed at low temperature, which agrees with our STM results at 78 K. The averaged intensity of the six equivalent fractional spots as a function of temperature is plotted in Fig. 2(g). By fitting the experimental data with a semi-phenomenological mean-field form

$$I(T) \propto \tanh\left(C\sqrt{\frac{T_c}{T} - 1}\right),$$

we got the CDW transition temperature $T_C \sim 192 \pm 2$ K. Our LEED results are similar to the recent report [23].

The LEED experiments reveal the averaged period of the distorted lattice structure of the CDW state in reciprocal space. To give the microscopic information on the lattice distortions as well as the charge-density modulation of the CDW state, we acquired STM images and d$I$/d$V$ maps on VTe$_2$ monolayer. The STM images and d$I$/d$V$ maps acquired in the same area from -0.20 V to +0.20 V are compared in Figs. 3(a) to 3(h). As indicated by the grey lines superposed on the images, each (4×4) unit cell can be divided into left and right half-cells (triangles pointing to the right and to the left, respectively). As shown by the d$I$/d$V$ maps from -0.20 V to -0.05 V, the right half-cells have higher local density of states (LDOS) than the left ones [right columns of Figs. 3(a) to 3(d)]. While the left half-cells show higher LDOS than the right ones in the d$I$/d$V$ maps from +0.05 V to +0.15 V [right columns of Figs. 3(f) to 3(h)], and at +0.20 V the LDOS in the two half-cells become comparable [right column of Fig. 3(e)]. The LDOS distribution revealed by d$I$/d$V$ maps is in accordance with the STS spectra measured at representative sites within the (4×4) cell: The right half-cell has higher LDOS intensity than the left one in negative bias range, while the left half-cell shows higher LDOS intensity than the right one in the positive bias range, and the corner site show the minimum LDOS intensity at both the negative and positive bias, as shown in Fig. 3(i).

In the STM images acquired at higher bias, |$V_{bias}$|>50 meV in our experiments, the contrasts are mainly contributed by the charge-density modulation, and thus the right half-cells show higher apparent height than the left ones from -0.20 V to -0.10 V [left columns of Figs. 3(a) to 3(c)], while the left half-cells are brighter than the right ones from +0.10 V to +0.20 V [left columns of Figs. 3(e) to 3(g)]. The STM images scanned at -0.05 V and +0.05 V show similar contrasts [left columns of Figs. 3(d) and 3(h)], and these STM images mainly reveal the atomic structure of the VTe$_2$ monolayer: The Te atom at the vertices of the unit cell protrudes from the surface, with a reduced in-plane distance to the surrounding 6 Te atoms, implying the in-plane distance between the Te atoms inside the left and right half-cells get increased.

As a consequence of the formation of CDW, a bandgap is expected at the Fermi level. Due to thermal broadening, such CDW gap cannot be observed at 78 K. Instead we measured STS on monolayer VTe$_2$ at 4.9 K. As shown in the up panel of Fig. 3(j), a dip feature is observed at the Fermi level. We further normalized the spectrum by dividing the spectrum with a cubic fitting of the background [12, 29], and extracted a CDW gap of 12 meV [bottom panel of Fig. 3(j)]. The dip feature is detected all over the VTe$_2$ monolayer at 4.9 K, see SM for more STS results [28].

The (4×4) lattice distortions and charge-density modulation in CDW state have broken the (1×1) translational symmetry in the normal state [cf. the STM image at 290 K, Fig. 1(c)]. Moreover, the real-space imaging provides more subtle information about symmetry breaking that cannot be observed by diffraction methods. By a careful investigation of the STM images and d$I$/d$V$ maps in Fig. 3, we found that the lattice distortions in the STM images and LDOS distributions in the STS maps lost the three-fold rotational symmetry as well as the mirror symmetry, which is not a typical behavior of the CDW states in TMD. In the (3×3) CDW of 2$H$-NbSe$_2$ monolayer [21], the (2×2) CDW of 1$T$-TiSe$_2$ monolayer [8], the (3×3) CDW of 1$H$-TaSe$_2$ monolayer [22], the (4×4) CDW of 1$T$-VSe$_2$ [12], and the ($\sqrt{13} \times \sqrt{13}$) CDW of 1$T$-TaS$_2$ [19], the CDW states preserve the three-fold rotational symmetry of the normal state since the lattice distortions in three high symmetry directions are identical.

In VTe$_2$ monolayer, as shown in Fig. 4(a), in addition to the (4×4) superstructure, the atomic lines along the dashed lines shows higher apparent height than their counterparts along the other two high-symmetry directions, and the overall lattice shows a 1D modulation. This 1D modulation in the lattice structure is evidently visualized in the Fourier transform of the image, as shown in Fig. 4(b). The spots perpendicular to the 1D modulation have lower intensity than those along the other two directions. To give a quantitative description of the 1D modulation of the lattice structure, we measured the in-plane distances of the six inner Te atoms within the half-cells [marked by colored circles in Fig. 4(a)] to the nearest corner Te atoms, as illustrated by the arrow in Fig. 4(a). The measured distances in various unit cells in Fig. 4(a) are plotted in Fig. 4(c). The Te atoms marked by blue, red and green circles have shorter distances than the other three species. In Fig. 4(d), we draw schematic models of the top-sublayer Te atoms in the normal state and the CDW state. The positions of the Te atoms in the CDW state are placed based on the STM image, and the displacements of the blue, red and green-colored Te atoms from the perfect (1×1) lattice are depicted by arrows. Clearly, the top-sublayer Te atoms break the three-fold rational and the mirror symmetries that exist in the normal state. Such 1D modulation is observed in other VTe$_2$ monolayers we have grown [28]. Since the LEED experiments detect large surface areas of the sample, the 1D modulation is smeared out over VTe$_2$ monolayers with different modulation directions. Therefore, such 1D modulation cannot be resolved in LEED experiments.

Considering the origin of the unique symmetry-breaking behavior of the CDW state in VTe$_2$ monolayer, we find that neither the conventional FSN mechanism [30] nor the q-dependent EPC interpretation [13, 31] can introduce such asymmetry component. In typical TMD systems, such as $2H$-NbSe$_2$, $2H$-TaSe$_2$, $1T$-TiS$_2$ and $1T$-TaS$_2$, the three-fold rotational symmetry in the normal state is always preserved in the CDW state, regardless of the origin of the CDW (either FSN or EPC still with some debates[13]). Consequently, for the monolayer VTe$_2$ we report here, either the FSN (or EPC) behaves differently with complete anisotropy to break the three-fold symmetry, or there is some other unknown mechanism breaking the equivalence of the symmetric directions. No matter what, monolayer VTe$_2$ provides a new unique platform to study the origin of CDWs in real 2D systems, which definitely needs deeper research in the future.

In summary, we have investigated the (4×4) CDW state of VTe$_2$ monolayer by LEED and STM. Our STM experiments resolved the lattice distortions and charge-density modulation of the (4×4) CDW in real space, and further unveiled a 1D structural modulation that breaks the three-fold rotational and mirror symmetries. Our work provides insights into the symmetry-breaking mechanisms of the CDW in TMD monolayers, and propose monolayer VTe$_2$ as a unique platform to study the origin of CDWs in real 2D systems.

Note: During the preparation of this manuscript, we noticed the appearance of a similar STM study [32] which has some overlap with the present work. Although the experimental STM results are consistent with each other, but the analyses and emphases of the two studies are quite different.

This work was supported by the National Key Research & Development Program of China (Nos. 2016YFA0300600 and 2016YFA0202300). X. Z. was supported by the National Natural Science Foundation of China (No. 11874404) and the Youth Innovation Promotion Association of Chinese Academy of Sciences (No. 2016008). J. G was partly supported by BAQIS Research Program No. Y18G09.

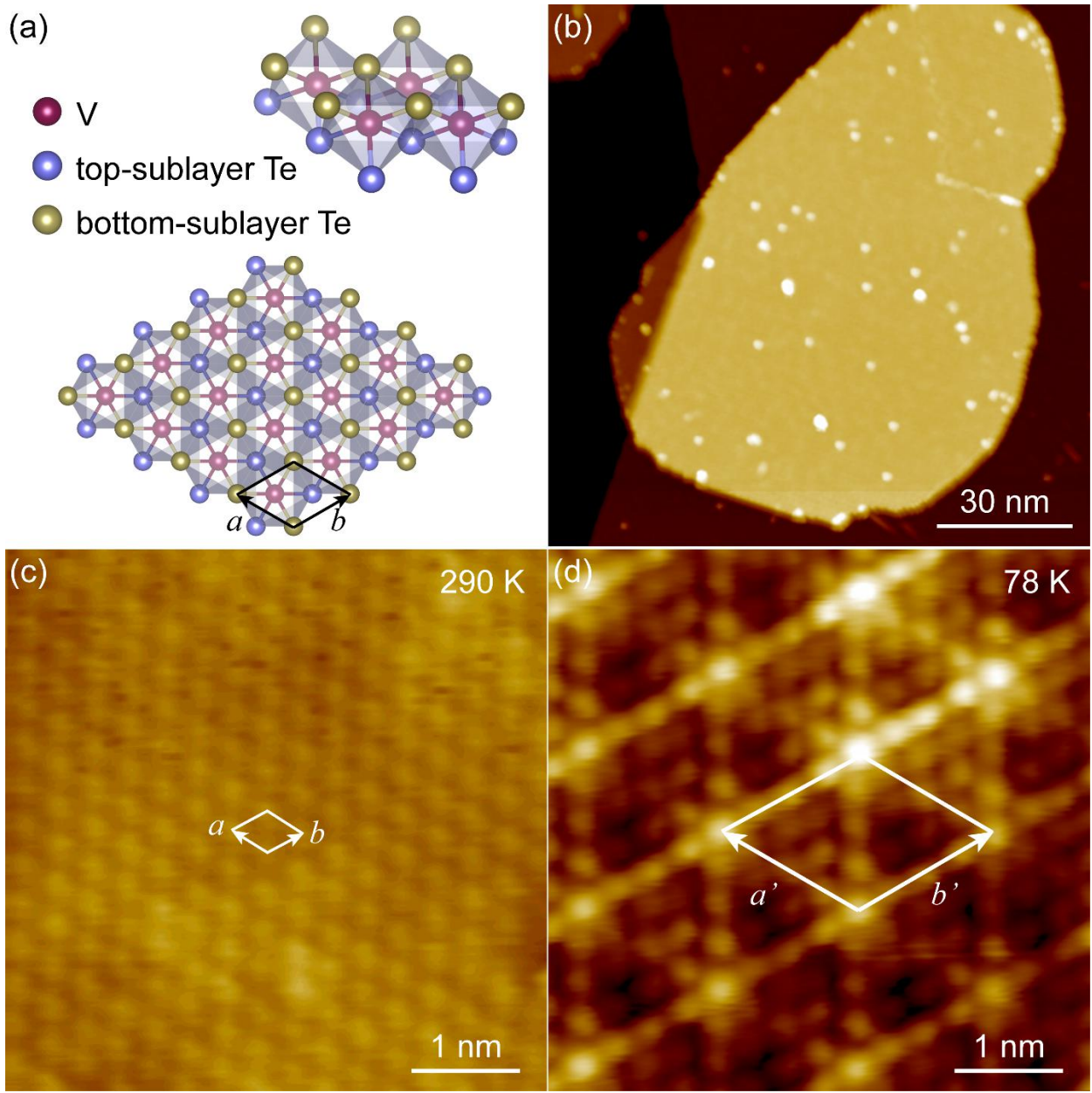

Figure 1 (a) Perspective (up) and top (bottom) views of the structural model of 1$T$-VTe$_2$ layer. (b) Large-scale STM image of a VTe$_2$ monolayer on graphene/SiC substrate (-2.0 V, 10 pA). (c) High-resolution STM image of VTe$_2$ monolayer scanned at 290 K (-0.75 V, 0.8 nA). (d) High-resolution STM image of VTe$_2$ monolayer scanned at 78 K (50 mV, 1 nA).

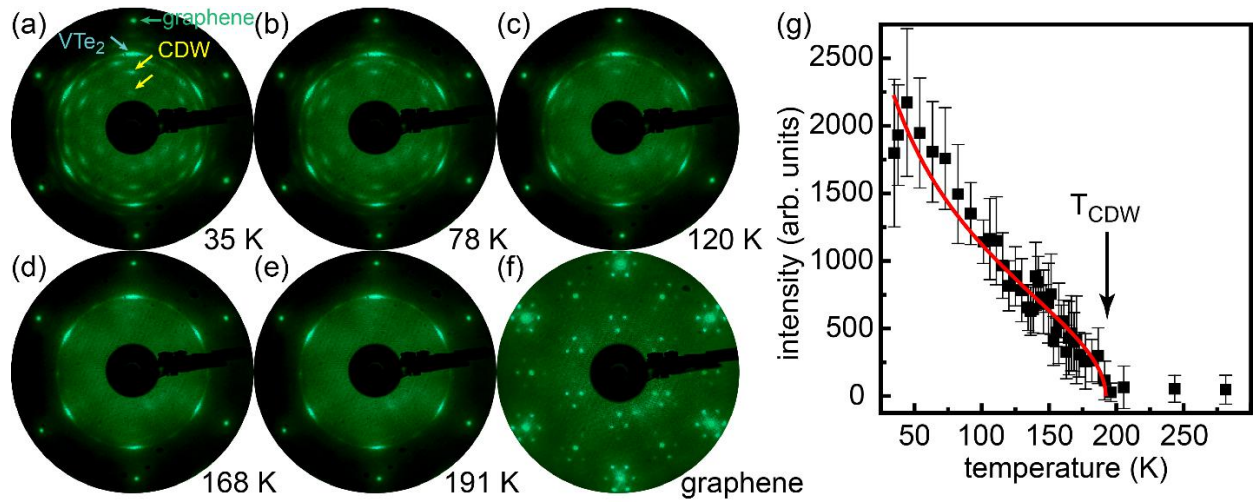

Figure 2 (a) to (e) LEED patterns measured on VTe$_2$ monolayer /graphene/SiC sample at indicated temperatures. The diffraction spots given by graphene, VTe$_2$ and CDW are indicated by arrows in (a). (f) LEED pattern measured on bare graphene/SiC substrate at 300 K. (g) The averaged intensity of the six equivalent fractional diffraction spots as a function of temperature are shown in squares, and the semiphenomenological mean-field fitting is plotted in red line.

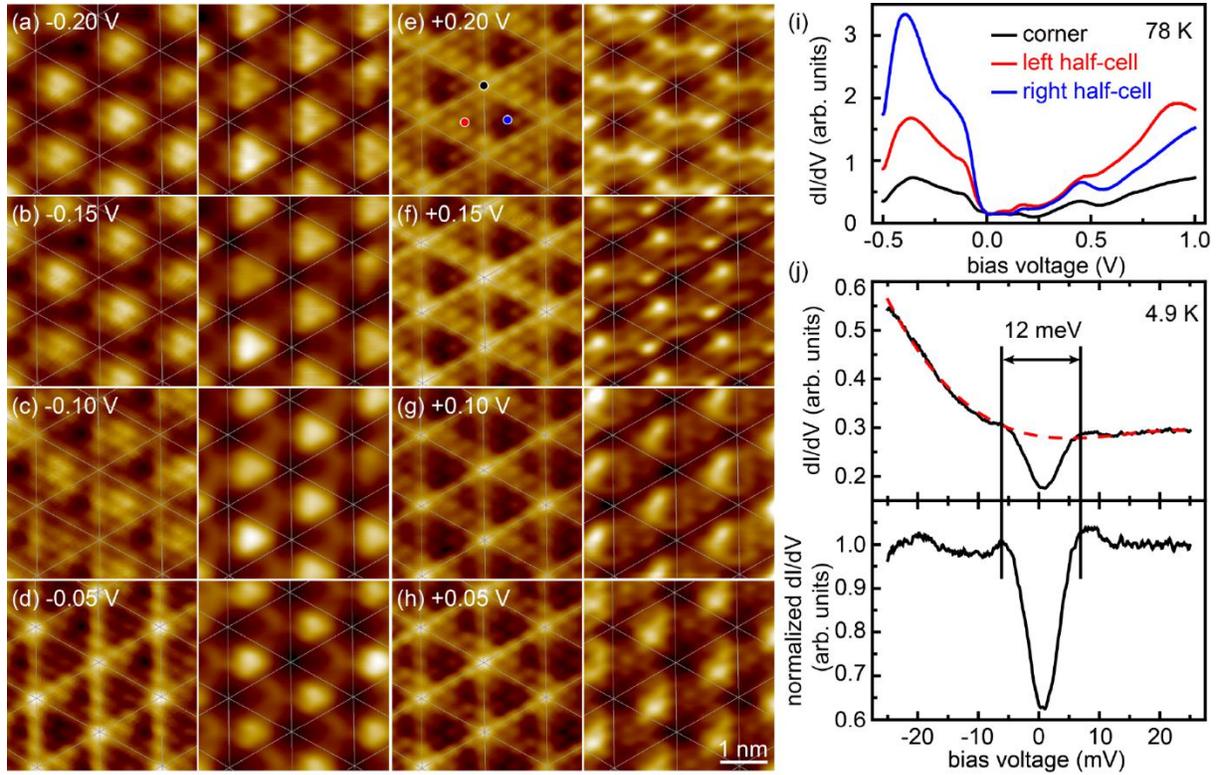

Figure 3 (a) to (h) STM images (left) and corresponding d$I$/d$V$ maps (right) at indicated bias voltages, all scanned at 78 K. (i) STS spectra measured at representative sites within the unit cell, as indicated in (e) following the color code. (j) Top: Typical STS spectrum measured on VTe$_2$ monolayer at 4.9 K. Bottom: Normalized spectrum by dividing the cubic fitting of the background. The cubic fitting of background is plotted in dashed line.

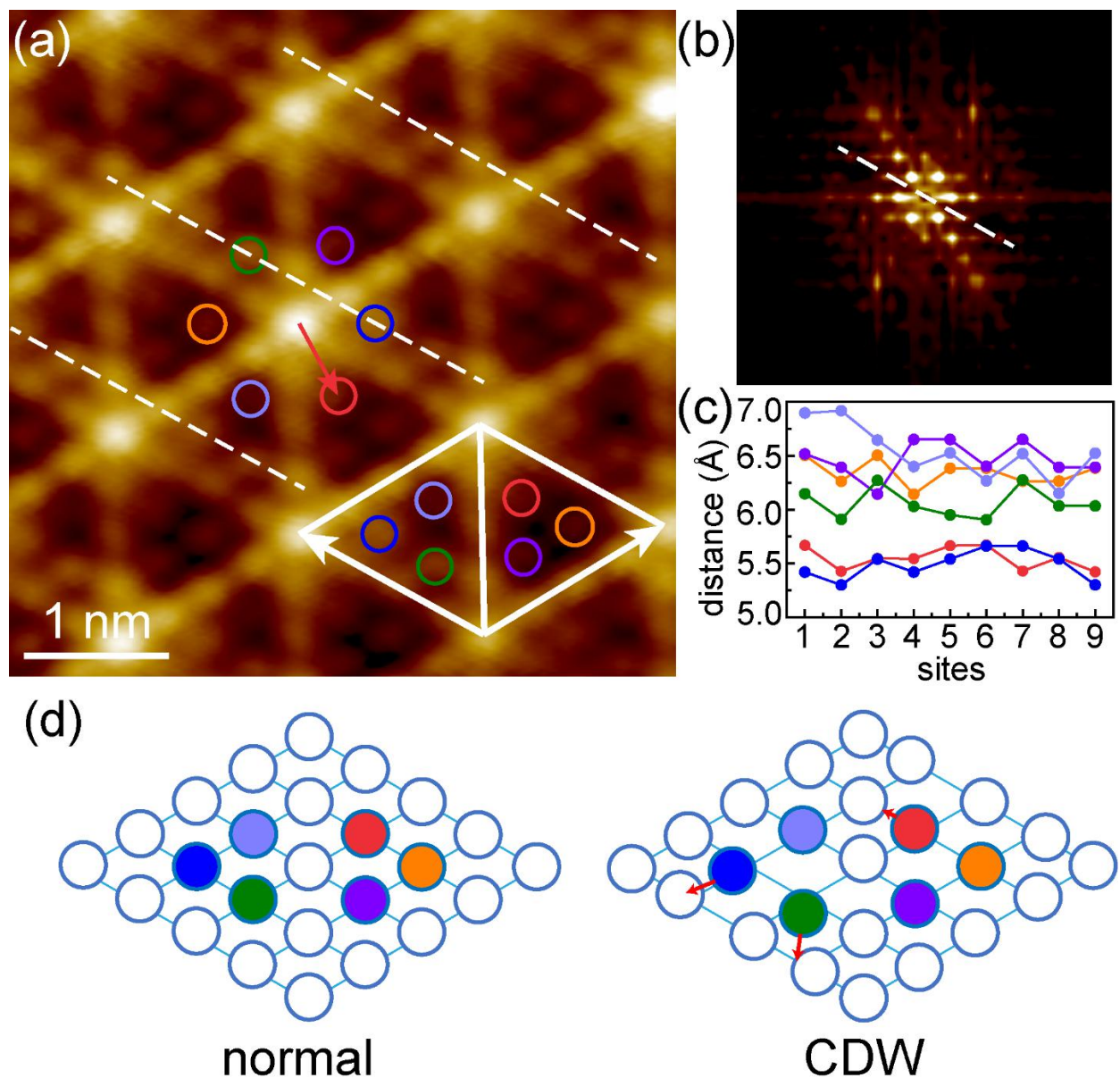

Figure 4 (a) STM image showing the 1D modulation of the (4×4) CDW state scanned at 78 K (20 mV, 0.5 nA). (b) Fourier transform of Fig. 4(a). The dashed lines in (a) and (b) indicate the 1D modulation. (c) Measured in-plane distances of the six inner Te atoms within the half-cells to the nearest corner Te atoms. The inner Te atoms are marked by colored circles in Fig. 4(a), and the distance of the Te atom marked by red to its nearest corner Te atom is indicated by the red arrow in Fig. 4(a). The distances of different inner atoms are plotted following the color code. (d) Schematic models of the top-sublayer Te atoms in normal state (left) and CDW state (right). The arrows indicate the displacements of the blue, red and green-colored Te atoms from the perfect (1×1) lattice.